Original Paper

# Adverse Childhood Experiences Ontology for Mental Health Surveillance, Research, and Evaluation: Advanced Knowledge Representation and Semantic Web Techniques

Jon Hael Brenas, PhD; Eun Kyong Shin, PhD; Arash Shaban-Nejad, MPH, PhD

Department of Pediatrics, Oak Ridge National Laboratory Center for Biomedical Informatics, University of Tennessee Health Science Center, Memphis, TN, United States

**Corresponding Author:**
Arash Shaban-Nejad, MPH, PhD
Department of Pediatrics
Oak Ridge National Laboratory Center for Biomedical Informatics
University of Tennessee Health Science Center
50 N Dunlap Street, R492
Memphis, TN, 38103
United States
Phone: 1 901 287 5836
Email: ashabann@uthsc.edu

## Abstract

**Background:** Adverse Childhood Experiences (ACEs), a set of negative events and processes that a person might encounter during childhood and adolescence, have been proven to be linked to increased risks of a multitude of negative health outcomes and conditions when children reach adulthood and beyond.

**Objective:** To better understand the relationship between ACEs and their relevant risk factors with associated health outcomes and to eventually design and implement preventive interventions, access to an integrated coherent dataset is needed. Therefore, we implemented a formal ontology as a resource to allow the mental health community to facilitate data integration and knowledge modeling and to improve ACEs' surveillance and research.

**Methods:** We use advanced knowledge representation and semantic Web tools and techniques to implement the ontology. The current implementation of the ontology is expressed in the description logic ALCRIQ(D), a sublogic of Web Ontology Language (OWL 2).

**Results:** The ACEs Ontology has been implemented and made available to the mental health community and the public via the BioPortal repository. Moreover, multiple use-case scenarios have been introduced to showcase and evaluate the usability of the ontology in action. The ontology was created to be used by major actors in the ACEs community with different applications, from the diagnosis of individuals and predicting potential negative outcomes that they might encounter to the prevention of ACEs in a population and designing interventions and policies.

**Conclusions:** The ACEs Ontology provides a uniform and reusable semantic network and an integrated knowledge structure for mental health practitioners and researchers to improve ACEs' surveillance and evaluation.

## Introduction

The study of Adverse Childhood Experiences (ACEs) and their consequences in terms of diseases and health risks has emerged during the past 20 years [1]. The US Centers for Disease Control and Prevention (CDC) demonstrated that ACEs are one of the root causes of several physical, social, cognitive, and emotional concerns [2], and affect nearly half of all US children under 18 years of age [3]. Direct indicators, including emotional, physical, and sexual abuse, at the individual level and indirect home environment indicators are commonly considered when studying ACEs. In general, ACEs are measured by two main experiences during upbringing and within the home environment: victimization experience and exposure to adversities in the family. Victimization includes emotional abuse, physical abuse, and sexual abuse. The exposure to adversities, often referred to as household dysfunction, is measured by family members' history of substance abuse, mental illness, domestic violence, incarceration, marital status, and financial difficulties leading to food deficiency. However, these factors influence health outcomes with different degrees and are closely linked to other adversities at the community level and other social determinants of health (SDH). Over the years, a connection between a higher number of ACEs and a variety of negative outcomes (eg, substance abuse [3], impaired memory [4], biological systems impairments [5], or criminal activities [6]) has been identified. ACEs are known to have complex and negative influences on individual health outcomes throughout the life course and have enduring negative influences on physiological outcomes as well as psychological functioning. ACEs do not only affect individuals but also have intergenerational impacts. Yet, many questions related to the ACEs' causal pathway and the effectiveness of existing interventions have remained unanswered: for example, deciding on whether the best intervention for a particular scenario is medical or a community-based approach [7].

The complexity of ACEs' causal pathway demands comprehensive and multidimensional data coordination. To design effective preventive interventions to reduce the burdens of ACEs, researchers and clinicians need access to a consistent knowledge-driven, evidence-based comprehensive analytic framework to study and monitor the causes of ACEs and their impacts on health (ie, obesity, mental health, and substance abuse), education (ie, cognitive developments, educational attainment, and graduation rates), and social dimensions (ie, placement in foster care and involvement with the justice system). The ideal ACEs knowledge-based system should be able to capture and identify ACEs indicators, detect individuals and groups at high risk of ACEs, integrate and validate ACEs health and social determinants, as well as exposure and genetic variations at individual and population levels. In order to improve the surveillance of ACEs, it is crucial to access a standard vocabulary that facilitates data collection, analysis, interpretation, and exchange of data between different parties and disciplines. An ontology is a standardized computational artifact that is used to capture, represent, and reason about the knowledge in the field. Ontologies capture knowledge by defining concepts, instances, relationships, and axioms. They increase the interoperability between different data sources and systems and improve the dissemination of data and knowledge across different disciplines. They also allow the use of semantic technologies to reveal new associations between the datasets and, therefore, discover new knowledge through logical inference. Ontologies are widely used in health and biomedicine and have made a substantial contribution to translational and clinical research as well as public and personalized health care [8].

The creation of the ACEs Ontology is rooted in recent efforts to study the effects of early experiences and social-environmental factors on children's development and life-course health. These efforts have been made by multiple actors from different disciplines at various levels, including international organizations (eg, the World Health Organization [9]), nongovernmental organizations (eg, the Foundation for Excellence in Mental Health Care [10] and the Center for Youth Wellness [11]), government offices (eg, the Tennessee Department of Children's Services [12] and the CDC [13]), volunteers that consolidate their efforts (eg, the ACEs Connection community [14] and the Cumbria Resilience Project [15]), and academic studies. All of these projects have a common goal, which is preventing occurrences of ACEs and providing assistance to victims; however, their results, studies, interventions, and tools are limited by the fact that they cannot easily be shared in a way that would improve the ability to repeat or reuse them elsewhere. The ACEs Ontology can bridge the gap between these efforts for efficient ACEs studies and surveillance.

While several ontologies exist in the domain of mental health, the ACEs Ontology is designed to deal with a carefully restricted scope, which is the study, prevention, and treatment of ACEs. We have included in the ontology some aspects that we deem important because they are tangentially connected to ACEs, such as potential health outcomes as well as their possible causes and aggravating conditions. Many other projects and resources exist that provide more detailed views of these adjacent domains; for example, the Children's Health Exposure Analysis Resource (CHEAR) project [16] deals with environment as well as measurable exposure and biological responses. Since environmental factors have impacts on the development of children and on the incidence of ACEs, we plan to use existing resources, such as the CHEAR ontology [17], in future versions of the ACEs Ontology. The same can be said of other adjacent domains, including the legal and penal system, psychiatry and psychology of children and adults, nutrition, and interventions.

In this paper, we present the ACEs Ontology as a formal reusable resource that can be used by the mental health research community to advance the surveillance and study of ACEs. The ACEs ontological structure provides a semantic backbone supporting the entire data model and hierarchy and supports logical inference and query answering. The ontology is used to form a consensus on the kind of data that is relevant and to create a common lexicon that makes it easier to share and reuse knowledge and information in the domain. The ACEs Ontology is an integral part of the Semantic Platform for Adverse Childhood Experiences Surveillance (SPACES) [18], which is a semantic recommender system currently under development. The SPACES system employs the contextual knowledge provided by the ACEs Ontology, along with a hybrid content- and context-based filtering approach to enable intelligent exploratory and explanatory analysis and informed decision making. In this paper, we will first explain how the ontology was built and how it will dynamically expand to meet the needs of the ever-evolving field of ACEs' surveillance and prevention. We will then show the usability of the ontology through a set of use-case scenarios. The paper will conclude with discussions of the findings and suggestions for future directions.

## Methods

To facilitate the integration of ACEs-related data from different data sources, the ontology reuses several existing ontological concepts and relations as much as possible, while introducing new components. In some cases, the ontology integrates multiple existing elements in order to represent new pieces of knowledge. The surveillance of ACEs is focused on monitoring, detection, and prevention of ACEs as well as studying their causality pathways and short- or long-term consequences. The knowledge required to describe these activities is linked to concepts and terminologies coming from a wide variety of domains, including medicine, legal justice, cognitive and personal behavior, and community responses. This means that building an integrated ontology and knowledge base that covers the whole range of relevant subjects is a tedious, time-consuming, and complex task.

To build the foundation for the ontology, we identified five different key pieces of knowledge in our scope of study. As shown in Figure 1, *Person* is used to representing human subjects. It encompasses the child as well as their family members and the rest of their social network. Felitti et al [1] provided a set of definitions and criteria for ACEs. ACEs may affect *Persons* at different levels of granularity (eg, divorce or separation of parents might affect all the children in a family, while being the victim of physical or sexual abuse might only affect one of them). *Social Determinants of Health* are upstream factors that can affect the *Person* and the exposure and manifestation of ACEs. SDH also act at different levels of granularity (eg, only members of a specific household are affected by the presence of mold at home, while the whole neighborhood is influenced, one way or another, by the lack of public transportation or safety). Proximity to a clinic or grocery store can be beneficial for all members of the community, while proximity to other institutions can be positive for only a portion of the community (eg, the presence of a strong religious community might positively influence its well-standing members, but may be seen negatively by people who do not belong). *Negative Health Outcomes* are the medical and surgical adversaries that can be caused or worsened by the ACEs (eg, higher risk of heart attack or cancer). *Intervention* represents the way that social workers, medical practitioners, or policy makers can mitigate or negate negative health effects. Interventions can be implemented through various channels (eg, medical, surgical, and legal). For instance, interventions can include heart surgery to treat a stroke or providing legal advice for people who fear that they might be evicted from their home. The ACEs Ontology contains more-detailed concepts to accurately describe these different overarching topics.

**Figure 1.** The top conceptual model for the Adverse Childhood Experiences (ACEs) Ontology, demonstrating the interactions between five major concepts: Person, ACEs, Social Determinants of Health, Interventions, and Negative Health Outcomes.

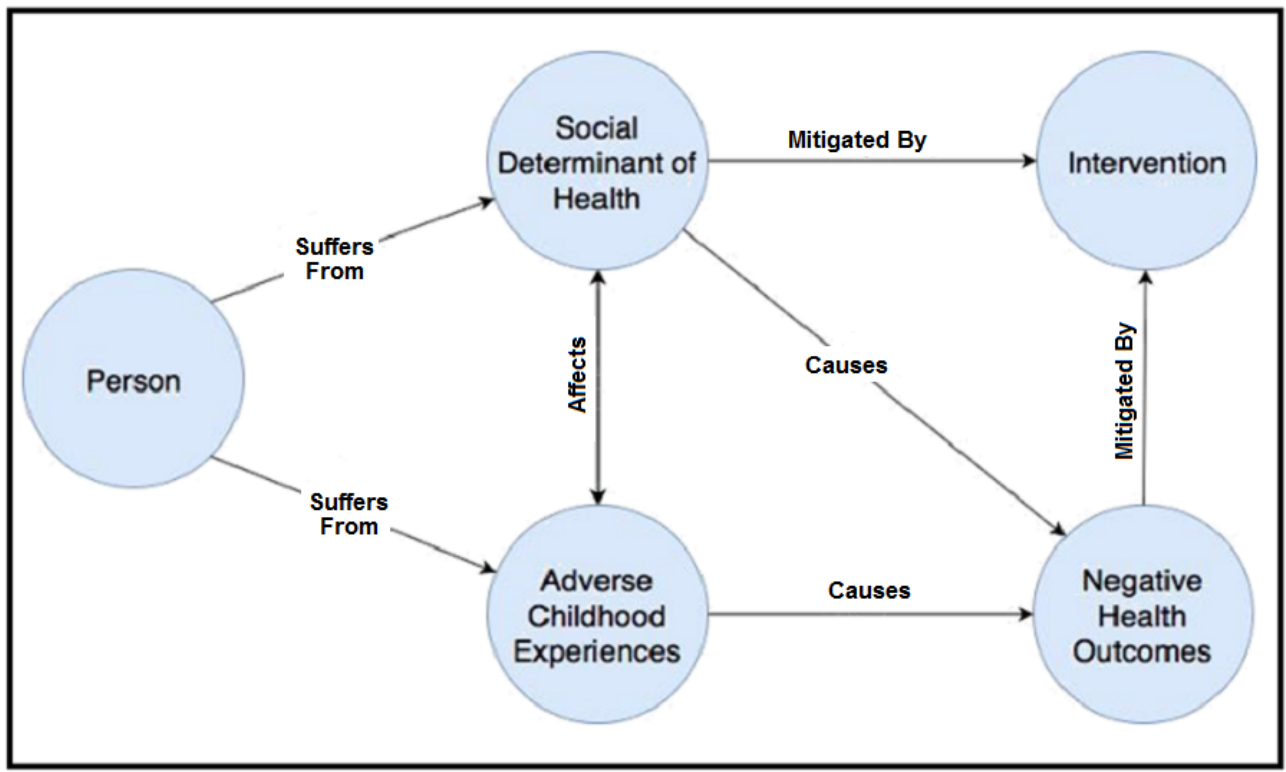

**Figure 2.** An abstract representation of some of the ontological concepts along with their relationships with the other thesauri in the field. SNOMED CT: Systematized Nomenclature of Medicine—Clinical Terms; ACEs: Adverse Childhood Experiences; NCIT: National Cancer Institute Thesaurus; MedDRA: Medical Dictionary for Regulatory Activities.

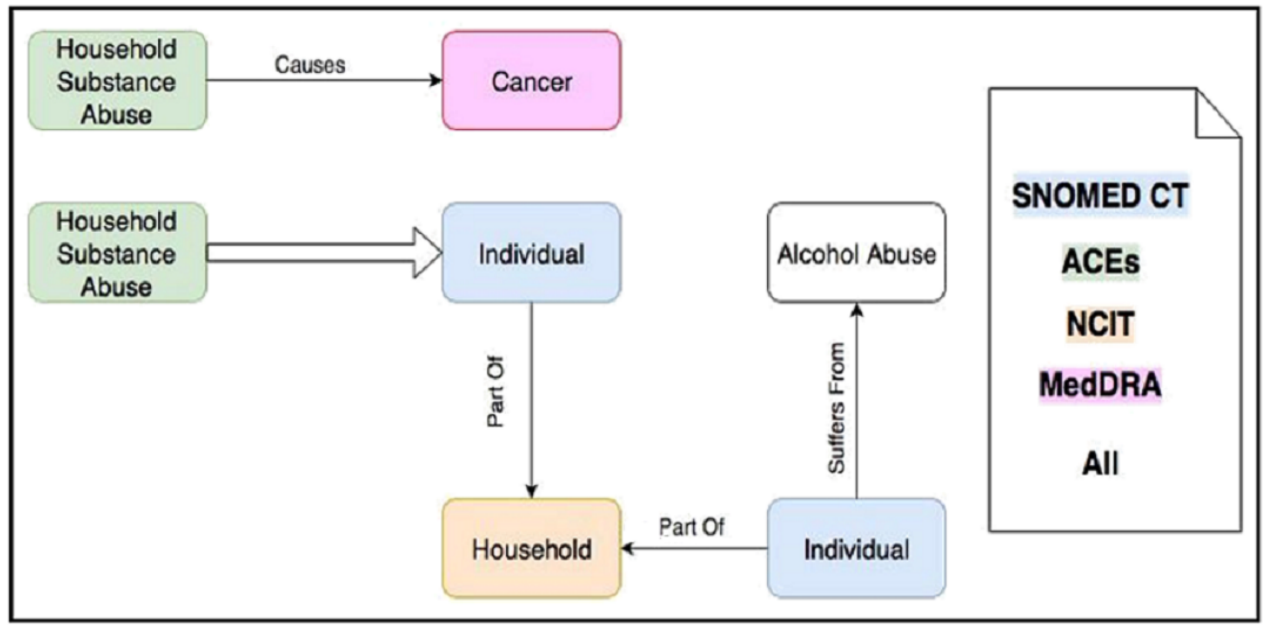

Figure 2 represents some examples of the major ontological elements; for instance, the ACE *Household Substance Abuse* affects *Individuals* who are part of the same *Household* as *Individuals* suffering from *Alcohol Abuse*— that *Household Substance Abuse* may itself be a causal factor for the development of *Cancer*. The concepts and relationships that appear in the ontology come from various sources in the field (eg, literature, databases, other ontologies, and domain experts). Many of the concepts that are central to ACEs (eg, abuse, violence, and mental illnesses and their medical consequences such as alcoholism, depression, obesity, and cancer) have been scattered among different existing biomedical ontologies and controlled vocabularies. For example, based on different needs, we imported a subset of the Systematized Nomenclature of Medicine—Clinical Terms (SNOMED CT) [19], the National Cancer Institute Thesaurus (NCIT) [20], and the Medical Dictionary for Regulatory Activities (MedDRA) [21] into the ACEs Ontology [22]. There is considerable overlap between the domains represented in these ontologies and, thus, it is not a surprise to find some repetitive concepts with similar or different definitions. In order to improve interoperability and reusability, we imported all the definitions and stated that they were equivalent, where applicable (ie, they describe the same concept).

## Results

### Overview

Figure 3 shows a fragment of the ACEs Ontology, including the partial concept and property hierarchies. The Web Ontology Language (OWL) format [23] is the de facto standard to express ontologies; for this reason, and for our ACEs Ontology to be freely reused by the community, the OWL format [23] of the ontology was made available via BioPortal [22] (see Figure 4).

An important decision when building the ontology was determining its overall expressivity. The more expressive the ontology, the higher the complexity of its associated reasoning tasks. In a similar way, increasing the complexity of an ontology makes it harder to maintain, as it becomes more difficult to make sure that no two inconsistent statements cohabit within. For this reason, and to encourage reusability of the ontologies, many biomedical ontologies focus almost exclusively on the taxonomical part of the ontology, which is defining inclusion relationships between various concepts (eg, *Verbal abuse* $\subseteq$ *Abuse*; for instance, verbal abuse is a subcategory of abuse) and avoiding the definition of more complex axioms (eg, *Verbal abuse* $\cong$ *Abuse* $\wedge \exists$ has Component. *Verbal*, for instance, verbal abuse is exactly the category of abuse that has a verbal component). In our ontology, we gradually define complex axioms as needed to capture more sophisticated key concepts necessary for ACEs' surveillance. General axioms, such as the ones shown in Figure 5, are used to express these complex concepts.

The current implementation of the ontology [22] is expressed in the description logic ALCRIQ (D), a sublogic of OWL 2 [23]. It is an expressive logic and, thus, the complexity of the decision problems is relatively high. In order to reduce the complexity, it is possible to avoid the use of datatypes, which are currently only used for dates and frequencies that might be handled as concepts; counting quantifiers, used to define the ACEs scores; or simplifying the role hierarchies. The ACEs Ontology currently contains 297 classes, 93 object properties, and 3 data properties. Most existing axioms are in the forms of class subsumption or equivalence.

**Figure 3.** A fragment of the Adverse Childhood Experiences (ACEs) Ontology representing partial class and property hierarchies. OWL: Web Ontology Language.

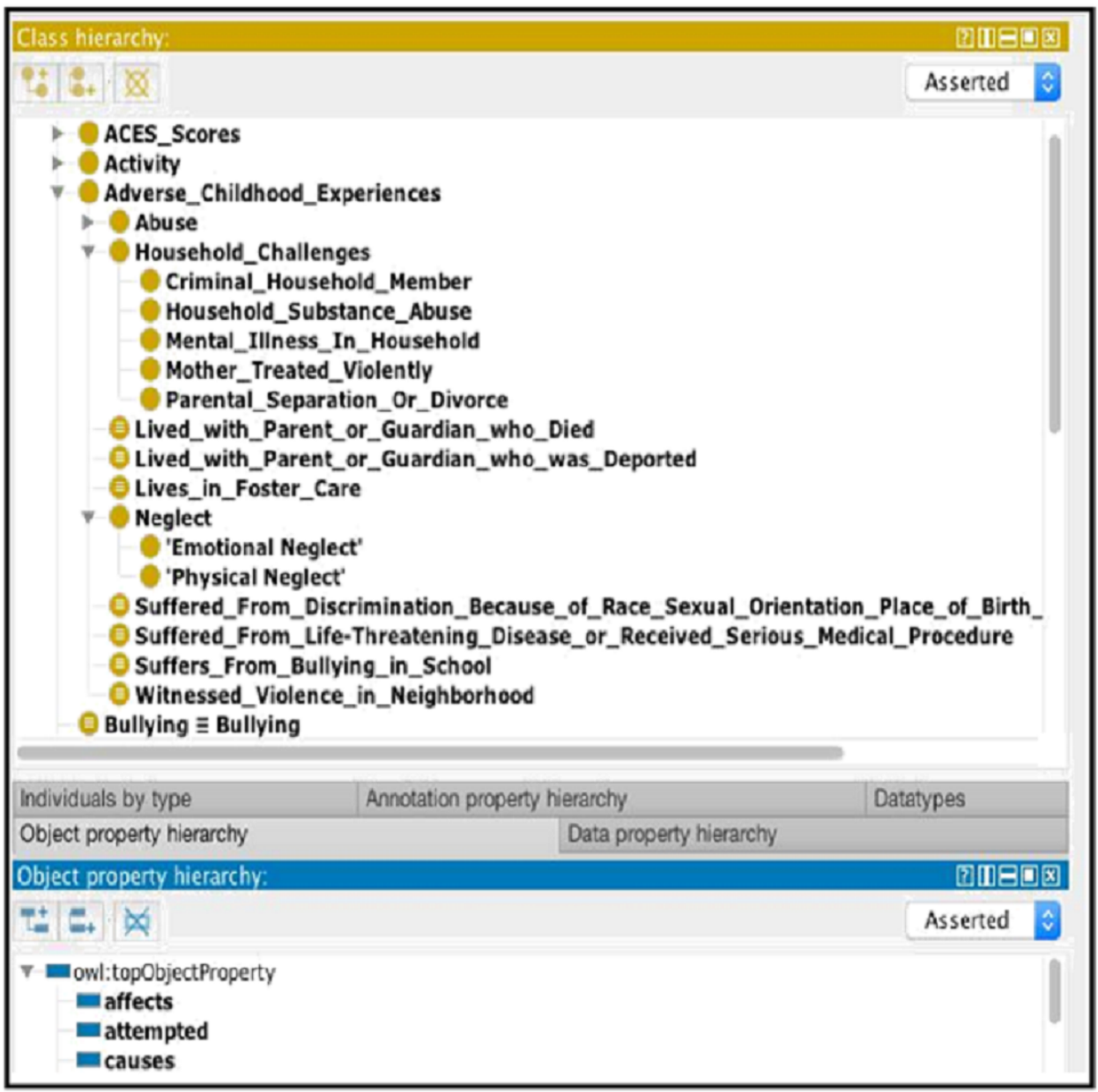

**Figure 4.** The Adverse Childhood Experiences (ACEs) Ontology on the National Center for Biomedical Ontology's (NCBO) BioPortal [22]. ACESO: Adverse Childhood Experiences Ontology; OWL: Web Ontology Language.

**Figure 5.** Some general axioms defined in the Adverse Childhood Experiences (ACEs) Ontology. OWL: Web Ontology Language.

In addition to the actual ontology, we also use semantic rules. In our ontology, some rules are used to enable data access. While the ontology provides a language to express the knowledge, the actual data itself is usually stored in data files or repositories that can use a different lexicon. Rules are employed to map the actual data to the ontology to enable knowledge-based querying and inference. Languages, such as Positional-Slotted, Object-Applicative (PSOA) Rule Markup Language (RuleML) [24], can be used to create those rules. Additionally, in our specific application to ACEs, some concepts require more expressivity than can be supported by OWL [21]. For instance, Felitti et al [1] define that one suffers from *Physical Abuse* if and only if a parent, stepparent, or adult living in her/his home throws something at her/him or pushes, grabs, slaps, or hits the person so hard that marks or injuries are caused. Someone suffered from *Physical Abuse* if, for example, the data contains a subgraph similar (ie, homomorphic) to the graph in Figure 6.

It is not possible to express exactly the existence of such a homomorphism in OWL. Several solutions can be considered. For instance, it is not possible to express that "i_p_h_t_r_i_i_t (inflicted_physical_harm_that_resulted_in_injuries_to) $\cong$ i_p_h_t_r_i_i (inflicted_physical_harm_that _resulted_in_injuries) o targets" but only that "i_p_h_t_r_i_i o targets $\subseteq$ i_p_h_t_r_i_i_t." As this representation does not allow us to fully express the knowledge and check its consistency, we can use an alternative solution by employing rules external to the ontology, such as "Physically Abused(x) $\rightarrow \exists$ y, z has parent(x, y) $\wedge$ i_p_h_t_r_i_i(y, z) $\wedge$ targets(z, x)," where x and y would span *Persons* and z would be an *Injury*. The right-hand side of the rule can be a logic formula, as in the example, or a query returning a Boolean value (eg, the query shown in Figure 7).

The third application of rules is to build a recommendation system. In that case, the left-hand side of a rule is a recommendation made to the user. Depending on the use case, the rule and the recommendation can take many different forms. For instance, if the ontology is used during an interview about ACEs with a child patient, the rule could be "If the patient's parents are separated or divorced, ask if they are feeling loved." These rules are particularly important in the context of ACEs detection due to the sensitive information (eg, mental illness, handicap, and criminal activity) that needs to be approached carefully.

Another key contribution of this kind of rule is to manage and allocate resources. For example, "If the patient suffers from emotional neglect, schedule an appointment with a child psychologist" (see Figure 8). Initially, these rules reflect a common sense reaction, but as the knowledge grows, new rules reflecting new findings can be added manually or procedurally.

By definition, rules are the result of causal reasoning (ie, if there is a match for the left-hand side, effect the right-hand side) and it is thus natural to connect the rules to causal diagrams. In the current state of this project, the rules are generated from common knowledge and currently established standard procedures and diagnostic processes. We employ causal diagrams [25], built from agreed-upon causality relationships that can be used to generate rules. As the goal of our recommender system is to go beyond what is currently known and practiced, we will use a combination of semantic and statistical inference models to infer the causal paths from the data and its use.

**Figure 6.** A piece of data representing someone (rectangular node) who was physically injured by a parent (round node). i_p_h_t_r_i_i_t: inflicted_physical_harm_that_resulted_in_injuries_to.

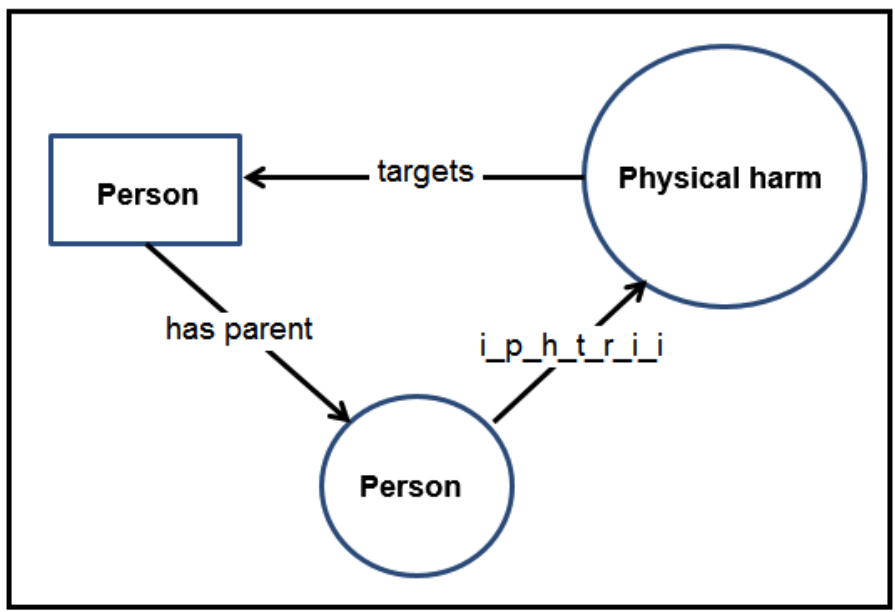

**Figure 7.** A SPARQL Protocol and Resource Description Framework (RDF) Query Language (SPARQL) query that could be the left-hand side of a rule instantiating the data with Physically Abused(x). i_p_h_t_r_i_i: inflicted_physical_harm_that _resulted_in_injuries.

```
ASK {
  ?child has_id x;
         has_parent ?parent.
  ?parent i_p_h_t_r_i_i ?physical_harm.
  ?physical_harm targets ?child.
}
```

**Figure 8.** A recommendation rule that schedules an appointment with a psychologist if a patient suffers from emotional abuse (Em_Ab).

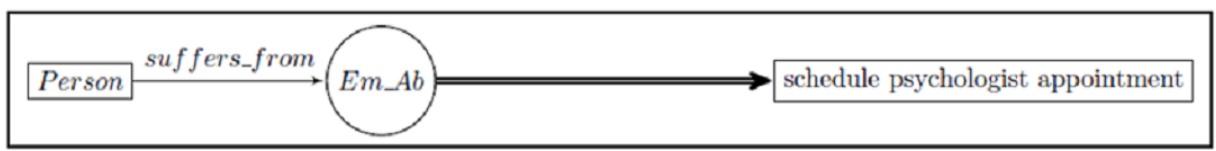

## Applications

We present three different use-case scenarios for the ACEs Ontology to showcase the role and significance of the ontology to improving ACEs' monitoring, detection, prevention, and management.

### Use Case 1: Clinical Setting

The first use case takes place inside a clinic where a nurse practitioner is interviewing a child patient and her parent. The parent gives personal identification data (ie, name, age, sex, and address) as well as a description of any potential symptoms. This information will be linked to other information that already exists about this patient in the electronic medical record. From the address, using a different data source that gathers data about social determinants of health, the ACEs Ontology helps to infer the patient's social economic status (eg, patient's access to public transportation, the neighborhood's safety and poverty rate, and residential proximity to schools and daycares). The medical practitioner can then use the collected information, as well as the inferred knowledge, to diagnose the illnesses or make a new hypothesis and continue to further investigate the case. Additionally, during the interview, the health practitioner will be able to ask questions relating to ACEs, for example, by slowly introducing household challenges (eg, "Are the parents divorced or separated?" or "Is a household member incarcerated?").

Depending on the answers, using the semantic links in the ontology, the medical practitioner will be able to ask more questions, recommended by the knowledge-based system, to reveal important clues and signs for detection of ACEs. The ontology is used both to allow access to different data sources that share important information and to foster reasoning that will facilitate the knowledge exchange between actors. In this scenario, the goal is mostly to increase the information and knowledge about the patient. Once the data is collected, however, it can be used to formulate a diagnosis. For instance, the ontology can be used to answer the question "Given that the patient has symptoms $S_0,..., S_n$ and an ACEs score of 4, which are the likely negative health outcomes to screen them for?"). An equivalent query in SPARQL Protocol and Resource Description Framework (RDF) Query Language (SPARQL) [26], a standard knowledge-based query language, is shown in Figure 9.

### Use Case 2: Public Health Policy Making

Policy makers and public health organizations can use the SPACES framework for knowledge-based population health surveillance [27] and to identify the causality pathway for the ACEs (eg, physical abuse of spouses and children) and design and implement interventions for remediating them. The semantic framework assists public health planners to compare different communities and programs with each other and finds the intervention (eg, reducing opioid addiction) that is best suited based on each community's needs and priorities. The ontology is used to access, share, and exchange data with other actors, stakeholders, and systems to allow for looking at the problem from various angles and with different granularities to formulate optimal responses. The ontology can then be used to answer questions such as "Given that resources $r_0,..., r_n$ are available, which intervention is most likely to reduce the prevalence of ACE a?" which can be expressed as the SPARQL query shown in Figure 10.

**Figure 9.** A SPARQL Protocol and Resource Description Framework (RDF) Query Language (SPARQL) query used to discover which negative health outcomes to screen a patient for. nho: negative health outcome; ACEs: Adverse Childhood Experiences.

```
SELECT ?nho\_name
WHERE {
      ?nho a negative_health_outcome;
           has_symptom ?s_0;
           ...
           has_symptom ?s_n.
      ?aces_score a aces_score_4;
                  increases_risk ?nho.
}
```

**Figure 10.** A SPARQL Protocol and Resource Description Framework (RDF) Query Language (SPARQL) query used to select an appropriate intervention.

```
SELECT ?name
WHERE {
        ?intervention has_name ?name;
                      has_ressource ?q_0;
                      ...
                      has_ressource ?q_n;
                      has_effect\_on\_a ?e.
FILTER(similarity(?q_0,r_0,..., ?q_n,r_n))
} ORDER BY ?e
```

*Use Case 3: Risk Stratification*

Another key use of the SPACES framework is in detecting potential problematic areas in order to intervene. As an example, because inflicting abuse is a criminal activity, collecting information on ACEs from the parents can yield less-reliable data than data about neighborhood conditions. It is thus easier to query the data about SDH that are prevalent in the neighborhood in which the patient is living and, thus, which ACEs are the most likely to be a risk. A typical semantic query for such cases can be formulated as "Given that the social determinants of health $S_0,..., S_n$ have been observed in the area, what are the ACEs to screen for?" Similarly, negative health outcomes are easier to track because they often result in hospital visits or medical interventions. The question "Given that the social determinants of health $S_0,..., S_n$ have been observed, and that the negative health outcomes $O_0,..., O_m$ have been frequent in the area, what are the ACEs to screen for?"

## Discussion

In this paper, we introduced an ontology for improving the surveillance of ACEs. The goal of the ontology is to provide a uniform structure to represent current and future studies on the causes and effects of, and ways to prevent and mitigate, ACEs. The benefits of using ontologies and semantic technologies have already been shown in several biomedical domains ranging from clinical surveillance [28] to global health surveillance [29]. After describing the architecture of the ontology and discussing the theoretical and engineering aspects involved, we presented multiple applications of the ontology to show how it could contribute to improving the surveillance of ACEs. The ontology was created to be used by major actors of the ACEs community with different applications, from the diagnosis of individuals and the potential negative outcomes that they encounter to the prevention of ACEs in a population and designing interventions and policies. We were able to evaluate the ACEs Ontology automatically using description logic reasoners to ensure its consistency and the satisfiability of its underlying semantic structure. In collaboration with domain experts, we also evaluated the quality of the ontology based on how well the ACEs Ontology aligned with the requirement criteria and standards defined in the project scope statement. Moreover, we assessed the usability of the ontology on the basis of its ability to answer the target queries. Furthermore, along with the further implementation of the SPACES platform, we also intend to design and develop a set of formal evaluation studies for in-depth assessment of the usability of the system and its impact on ACEs' surveillance and decision making.

The ontology is constantly growing with the advances in knowledge by adding new relevant concepts and relationships as well as new axioms. It is possible for some applications to import only the taxonomy to represent data in a standardized way, while others might use constructs such as transitivity for more complex reasoning. However, the benefits of the ontology are limited by the quality of the data available. Currently, most existing studies deal with adults for whom the negative outcomes are apparent, while more transversal studies of the impact of ACEs on the development of children are lacking [30]. We are currently developing a tool that would use the ontology and semantic inference alongside the statistical inference to assist us in finding patterns in the structured and nonstructured datasets to answer exploratory and explanatory questions.

## Acknowledgments

We would like to thank Dr Robert L Davis, Dr Jonathan A McCullers, Dr Sandra R Arnold, and the entire team at the Family Resilience Initiative at Le Bonheur Children's Hospital, Memphis, Tennessee, for their support and insights. This research was supported by the Memphis Research Consortium.

## Conflicts of Interest

None declared.

## Abbreviations

**ACE:** Adverse Childhood Experience
**ACESO:** Adverse Childhood Experiences Ontology
**CDC:** US Centers for Disease Control and Prevention
**CHEAR:** Children's Health Exposure Analysis Resource
**Em_Ab:** emotional abuse
**i_p_h_t_r_i_i:** inflicted_physical_harm_that _resulted_in_injuries
**i_p_h_t_r_i_i_t:** inflicted_physical_harm_that_resulted_in_injuries_to
**MedDRA:** Medical Dictionary for Regulatory Activities
**NCBO:** National Center for Biomedical Ontology
**NCIT:** National Cancer Institute Thesaurus
**nho:** negative health outcome
**OWL:** Web Ontology Language
**PSOA:** Positional-Slotted, Object-Applicative
**RDF:** Resource Description Framework
**RuleML:** Rule Markup Language
**SDH:** social determinants of health
**SNOMED CT:** Systematized Nomenclature of Medicine—Clinical Terms
**SPACES:** Semantic Platform for Adverse Childhood Experiences Surveillance
**SPARQL:** SPARQL Protocol and RDF Query Language